\documentstyle[12pt]{article}

\begin{document}

\begin{center}

{\large{\bf Self-Similar Law of Energy Release before Materials
Fracture} \\ [5mm]

V.I. Yukalov\footnote{The corresponding author}} \\ [2mm]
{\it
Bogolubov Laboratory of Theoretical Physics \\
Joint Institute for Nuclear Research, Dubna 141980, Russia} \\ [5mm]

{\large A. Moura and H. Nechad} \\ [2mm]
{\it
GEMPPM, UMR CNRS 5510 (B\^at. B. Pascal) \\
Institut National des Sciences Appliquees de Lyon \\
20 av. A. Einstein, 69 621 Villeurbanne, France} \\

\end{center}

\vskip 2cm

\begin{abstract}

A general law of energy release is derived for stressed heterogeneous
materials, being valid from the starting moment of loading till the
moment of materials fracture. This law is obtained by employing the
extrapolation technique of the self-similar approximation theory.
Experiments are accomplished measuring the energy release for industrial
composite samples. The derived analytical law is confronted with these
experimental data as well as with the known experimental data for other
materials.

\end{abstract}

\vskip 1cm


\newpage

\section{Introduction}

The study of global failure of stressed heterogeneous materials is of
great importance for the possibility to control the state of health of
complex structures in industry. The majority of industrial materials are
composites that, despite their regular appearance at a macroscopic level,
intrinsically possess numerous mesoscopic defects, i.e. defects that are
neither microscopic nor macroscopic, but which are intermediate between
the latter, say at the scale of fibres. These intrinsic imperfections
give rise to high local stress concentrations when such composite
materials are strongly loaded. Contrary to perfect crystals, whose
failure occurs suddenly, due to the cleavage effect with unstable crack
propagation, the failure of nonperfect materials develops as a multistep
procedure through a succession of local events corresponding to diffuse
damage. The progressive growth of the local damages, such as crack
nucleation, fibre-matrix debonding and rupture, and matrix separations,
culminate in the catastrophic rise of a global crack implying the
materials fracture. The gradual process of the defect accumulation,
developing before the materials failure, is accompanied by acoustic
emission characterizing the energy released from the stressed sample
(Broberg, 1999).

To understand the general law of the energy release from composite
materials would be of high importance for many applications and could be
employed for predicting the fracture of loaded industrial structures.
Unfortunately, the general form of such a law for the energy release,
describing the behaviour of the latter from the very beginning of loading
till the final moment of fracture, is not known. It is the aim of the
present paper to derive such a general law of energy release and to
confront it with experimental data.

\section{Physical picture of rupture}

We consider the failure of stressed composite materials, which occurs
through the gradual accumulation of local mesoscopic defects, accompanied
by a measurable energy release. The physical picture of the process,
developing in the {\it close vicinity} of materials fracture is more or
less generally accepted. The fracturing of stressed composite materials is
somewhat analogous to a critical phenomenon happening at a second-order
phase transition (Herrmann and Roux, 1990). The moment of rupture is similar
to a critical point, so that the fracturing process can be described by a
renormalization-group scheme (Anifrani {\it et al}., 1995). In the vicinity
of the critical point of rupture, there exists a critical region (Lamagh\'ere
{\it et al}., 1996; Sornette and Andersen, 1998), where the energy release
can be characterized by a power law decorated by log-periodic oscillations
(Sornette, 1998). Such oscillations are related to complex exponents that
appear in renormalization-group solutions for critical phenomena (Nauenberg,
1975).

The multistep character of arising mesoscopic defects, reflecting the
discrete nature of the fracturing process, makes it possible to compare
the diffusion of damages with the diffusion on random lattices (Bernasconi
and Schneider, 1983; Dietrich and Sornette, 1998). This analogy has been
the starting idea for extensive numerical simulations applied to the
lattice models of fracture (Sahimi and Arbabi, 1996), which predicted
the existence of scaling laws in the vicinity of fracture. With regard to
real heterogeneous materials, it is important to emphasize two principal
points:

\vskip 2mm

First, we keep in mind the composite materials with a random spatial
distribution of defects. Because of this spatial randomness, the growth
of damages under a uniform loading happens randomly across the volume of
the sample. A sufficiently random distribution of defects is naturally
related to a high degree of their concentration. The stronger is the
degree of imperfection, the more accurate are the scaling laws in the
vicinity of rupture and the closer is the similarity of the latter with a
critical phenomenon (Johansen and Sornette, 2000).

\vskip 2mm

Second, not only the growth of numerous damages happens randomly in
space, but the local growth of each of them randomly stops due to the
internal structure of matter. For instance, in a random porous material,
the growth of a crack ends as soon as it encounters a pore. In the case
of composite materials, there are three main types of damages: matrix
cracking, fibre rupture, and matrix-fibre debonding. These damages grow
simultaneously and, being intermixed, they create a spatially random
distribution of defects. It is worth recalling that even in brittle
composite materials, where the fracture occurs rather quickly, the dynamic
stress intensity factor at the crack tip dramatically reduces as soon as
the crack encounters a fibre, which leads to the crack arrest (Broberg,
1999).

\vskip 2mm

In this way, in a stressed composite material, with the described
properties, there arises the proliferation of randomly distributed defects.
Close to the point of rupture, the concentration of defects is so high
that they form a random self-organized structure. The vicinity of rupture
is analogous to the critical region, and the global materials failure
corresponds to a critical point. Then, being based on renormalization-group
arguments, one comes to the conclusion that the energy release close to
the point of rupture follows a power law decorated by log-periodic
oscillations (Sornette, 1998).

Although the asymptotic behaviour of the energy release in the close
vicinity of materials rupture can be written explicitly, its general law
in the whole region, starting from the absence of any load till the
moment of rupture, is not known. But, as is clear, the knowledge of this
general law would be of great importance. For instance, knowing such a
law, one could, on the basis of early observations, predict the following
behaviour of a stressed sample and even forecast the moment of its fracture.

In the recent communication (Moura and Yukalov, 2002), we suggested a
procedure for deriving the general law of energy release by employing the
self-similar approximation theory (Yukalov {\it et al}., 1990, 1991, 1992,
1997-a-b, 1998-a-b). However, for this purpose, we considered only the
simplest one-step variant of the procedure, which, though allowing for
the extrapolation outside the critical region, could not provide a very
accurate description far from the latter region, especially at the very
beginning of the loading process. In the present paper, we invoke a more
elaborate variant of the self-similar approximation theory for deriving a
really general law of energy release for stressed composite materials,
the law that would be valid starting from zero load till the moment of
fracture.

Before passing to technical details it is important to clarify several
points:

First of all, we consider here the energy release as the main characteristic
describing the process of fracturing. This is because the energy release
is the quantity that can be {\it directly measured} in acoustic emission
experiments. In principle, one could invoke some other characteristics
describing the evolution of the material structure. Thus, in the
classical models of continuum damage mechanics, one considers a quantity
playing the role of a material-structure parameter (Kachanov, 1986). The
structure parameter can be rather convenient for describing the spatial
state of continuous medium. It could also be useful at the very beginning
of the fracturing process in heterogeneous non-plastic materials.
However, the continuous mechanism is no longer valid when the studied
material becomes effectively discontinuous at the advanced stage of
fracturing, owing to high concentration of random defects. Even if it
would be possible to attribute a kind of structural order parameter to
composite materials, as those we consider, we prefer to deal with the
energy release. This is not because other characteristics are not
admissible, but just because the energy release is a directly measurable
quantity, hence, it is the most convenient for comparison with experiments.

The oscillatory behaviour of the energy release in the vicinity of
fracture can be presented in several slightly different ways. We opt here
for the so-called log-periodic form suggested and studied in several
papers (Anifrani {\it et al.}, 1995; Johansen and Sornette, 2000). This
log-periodic form can be justified by explicit mechanical models (Dietrich
and Sornette, 1998; Sornette and Andersen, 1998). The models of rupture
yielding the log-periodicity, as well as the physical interpretation of
the related parameters, have been analysed in the recent review by Sornette
(1998), because of which there is no reason to repeat them here.

One should not forget that the log-periodic presentation models the
behaviour of the energy release only in the vicinity of rupture, thus,
being invalid far from the fracture point. In order to extrapolate this
log-periodic form to the whole process of loading, starting from the very
beginning till the moment of fracture, one needs to employ some
extrapolation techniques. For this purpose, we use the {\it self-similar
approximation theory} (Yukalov, 1990, 1991, 1992; Yukalov {\it at al}.,
1997-a-b, 1998-a-b). To make this paper more self-consistent, we describe
the main ideas of this theory in Appendix.

It is worth emphasizing that the employed self-similar extrapolation
technique is quite general and can be used for extrapolating a variety
of functions, but not solely those having the sinusoidal log-periodic
form. For instance, there exists a wide class of parametric homogeneous
functions, including as a particular case the sinusoidal log-periodic
functions (Borodich, 1997, 1998-a-b; Borodich and Galanov, 2002). If the
behaviour of the energy release in the close vicinity of fracture could
be presented through such a parametric homogeneous function, one could
also extrapolate this behaviour by employing the self-similar
approximation theory. Our paper provides a general method of such an
extrapolation. The usage of the sinusoidal log-periodic presentation is
not compulsory, but just it is motivated by several known models of
fracture (Sornette, 1998).

\section{Derivation of general law}

The technique of the derivation is as follows. We start from the critical
region of rupture, where the energy release is known to be asymptotically
presented by a power law decorated by log-periodic oscillations
(Sornette, 1998). Then, we extrapolate this asymptotic form outside the
critical region by means of the self-similar approximation theory
(Yukalov {\it et al}., 1990, 1991, 1992, 1997-a-b, 1998-a-b). This theory
provides a general tool for extrapolating asymptotic expansions to the
whole region of variables, and it has been successfully applied to describing
critical phenomena as well as to time-series analysis (Yukalov {\it et al}.,
1999, 2000, 2001). All technical details of the self-similar approximation
theory can be found in the cited references. For the convenience of the
reader, the basic ideas of this approach are sketched in Appendix.

To correctly formulate the notion of the group self-similarity on the
manifold of approximations, we have, first of all, to present the sought
physical quantity in a scale-invariant form. To this end, we need, first,
to choose a convenient dimensionless variable for the energy release. The
latter can be considered as a function of the uniformly and isotropically
imposed pressure $p$ changing from $p=0$ to the critical pressure of
rupture $p_c$. In the case of an anisotropic load, one can measure the
energy release as a function of elongation $l$, with $l$ varying from
zero to a critical elongation $l_c$, when the stressed material breaks.
One may also treat the energy release as a function of time varying from
the initial moment $t=0$, when there is no load, till the critical moment
$t_c$, when the loaded sample is fractured. In each of such cases, we can
introduce a relative dimensionless variable defined by one of the
following forms:
\begin{equation}
\label{1}
x \equiv \left \{ \frac{p_c-p}{p_c}\; , \frac{l_c-l}{l_c}\; ,
\frac{t_c-t}{t_c} \right \} \; ,
\end{equation}
with $x$ given in the domain $[0,1]$. Then, if $E(x)$ is a cumulative
energy release, changing from $E(1)=0$ at the start of the loading
process to $E(0)=E_c$ at the moment of rupture, we define the reduced
energy release
\begin{equation}
\label{2}
f(x) \equiv \frac{E(x)}{E(0)} \; .
\end{equation}
By this definition, function (2) satisfies the boundary conditions
\begin{equation}
\label{3}
f(0) =1 \; , \qquad f(1) = 0 \; .
\end{equation}
There are two other physical conditions
\begin{equation}
\label{4}
f(x) \geq 0 \; , \qquad f'(x) \leq 0 \qquad (0\leq x\leq 1) \; ,
\end{equation}
requiring that both the cumulative energy release and the energy release
rate be positive. Sometimes, it is useful to invoke the integral
normalization condition
\begin{equation}
\label{5}
\int_0^1 f(x)\; dx = \Phi \; ,
\end{equation}
in which the value $\Phi$ is prescribed by the considered experiment.

In the asymptotic vicinity of the critical point $x=0$, the dimensionless
energy release (2) follows, as is discussed in the previous section, the
power law decorated by log-periodic oscillations, which reads
\begin{equation}
\label{6}
f(x) \simeq 1 + a_1\; x^\alpha +
a_2\; x^\alpha \cos(\omega\ln x +\varphi) \; ,
\end{equation}
as $x\rightarrow 0$, where $a_1,\; a_2,\; \omega,\; \alpha$, and $\varphi$
are the parameters characterizing the studied material, because of which
they can be called the {\it material parameters}. The most physically
important here is the parameter $\omega$ characterizing  the frequency of
the log-periodic oscillations, while $a_2$ describes their amplitude. As
is explained above, form (6) follows from some models of fracture
(Sornette, 1998).

Function (6) can be treated as the real part of the complex expression
\begin{equation}
\label{7}
F(x) \simeq 1 + A_1\; x^\alpha + A_2\; x^{\alpha+i\omega} \; ,
\end{equation}
in which $x\rightarrow 0$ and the coefficients are
$$
A_1 \equiv a_1 \; , \qquad A_2 \equiv a_2\; e^{i\varphi} \; .
$$
Expansion (7) can be renormalized employing the self-similar approximation
theory (Yukalov {\it et al}., 1990, 1991, 1992, 1997-a-b, 1998-a-b). For
this purpose, each term in Eq. (7) is considered as a correction to the
sum of the previous terms. The self-similar renormalization of Eq. (7),
based on the self-similar exponential approximants (Yukalov and Gluzman,
1998-a), results in
\begin{equation}
\label{8}
F^*(x) =\exp\left ( c_1x^\alpha\; \exp\left ( c_2x^{i\omega}\right )
\right ) \; ,
\end{equation}
where
$$
c_1 \equiv A_1\tau_1 \; , \qquad c_2 \equiv \frac{A_2}{A_1}\; \tau_1
$$
are the so-called controllers, with $\tau_1$ and $\tau_2$ being the
step-control functions to be defined from optimization conditions. Since
the sought function $f(x)$ is the real part of Eq. (7), then the real
part of Eq. (8) provides us with the self-similar approximant
\begin{equation}
\label{9}
f^*(x) ={\rm Re}\; F^*(x) \; .
\end{equation}
Taking the real part of Eq. (8), it is convenient to pass to the notation
$$
c_1 \equiv c \; , \qquad c_2 \equiv a+i\; b \; ,
$$
in which all parameters $a,\; b$, and $c$ are real. In this way, we obtain
\begin{equation}
\label{10}
f^*(x) =\cos\left [ cx^\alpha\sin g(x)\; \exp h(x)\right ] \;
\exp\left [ cx^\alpha\cos g(x)\; \exp h(x) \right ] \; ,
\end{equation}
where we introduce the functions
$$
g(x) \equiv a\sin(\omega\;\ln x) + b\cos (\omega\; \ln x) \; , \qquad
h(x) \equiv a\cos(\omega\;\ln x) - b\sin(\omega\; \ln x) \; .
$$

Expression (10) is the sought general law for the energy release,
extrapolating the asymptotic expansion (6) to the whole region of the
variable $x\in[0,1]$. Because $f^*(0)=1$, the first of the boundary
conditions (3) is automatically valid. The second of these boundary
conditions requires the validity of the equation
$$
c\; e^a \sin b = \frac{\pi}{2} + \pi n \; ,
$$
in which $n$ is an integer. Since the latter is arbitrary, this condition
does not impose a strict constraint, so that in what follows five material
parameters $a,\;b,\;c,\;\alpha$, and $\omega$ can be treated as independent.
The values of these parameters can be found by comparing the law (10) with
experiments. The physical meaning of these parameters is clear from
expression (10): the parameter $\omega$ describes the log-periodic
frequency of oscillations; the parameters $a$ and $b$, their amplitude;
and the parameters $c$ and $\alpha$ characterize the rate of the overall
increase of the released energy.

\section{Comparison with experimental data}

We accomplished experiments studying the behaviour of the cumulative
energy release as a function of the dimensionless parameter (1). The
studied samples were the glass-polyester composite plates of the sizes
$2\times 14\times 120$ mm. Each plate contained $75\%$ of unidirectional
fibres of diameter $20 \;\mu$m. At the initial moment, a plate was
subject to a steady tensile stress of $650$ N at an angle of $27^o$ to
the fibre direction, which forced the material to creep. Varying the angle
does not result in the principal change of the process. During the
creeping test, both acoustic emission and elongation were recorded up to
the global fracture. Piezoelectric transducers recording the Lamb waves,
i.e. guided waves in the plate, were employed. Electric signals,
reflecting the energy release rate caused by sudden local damages, were
used for calculating the cumulative energy release. The stressed sample
was inside an oven supporting a constant temperature of $60^o$C. The
general experimental setup is shown in Fig. 1 and more details are
described by Moura {\it et al}. (2002). Particular characteristics of
these experiments could be varied, which, however, would not essentially
influence the final results of measurements, provided these results are
presented in the scale-invariant dimensionless form (2). A typical
behaviour of the dimensionless cumulative energy release (2) versus the
dimensionless elongation (1), observed in our experiments, is shown in
Fig. 2, where it is compared with the law (10); the fitting material
parameters being $a=0.084$, $b=-0.364$, $c=-4.05$, $\alpha=0.613$, and
$\omega=1.49$.

Note that the derivative of $f^*(x)$, corresponding to the energy release
rate, diverges at the point of global failure, which confirms the analogy
of the latter with a critical point. We may also define the critical index
$$
\Delta \equiv \lim_{x\rightarrow+0}\;
\frac{\ln|f'(x)|}{\ln x} \; ,
$$
for which we find $\Delta=\alpha-1=-0.387$.

Among other known experiments on the energy release measurements, the
most reliable are those accomplished by Anifrani {\it et al}. (1995) who
recorded the acoustic emission during the uniform pressurization of
spherical tanks of kevlar and carbon fibres impregnated with resin. This
composite material was wrapped up around a thin metallic liner (steal or
titanium). The samples were fabricated by A\'erospatial-Matra Inc.. The
general experimental methodology was similar to ours, with the difference
that the dimensionless variable (1) represented the change of pressure.
In Fig. 3, we compare the law (10) with typical acoustic-emission
measurements accomplished by Anifrani {\it et al}. (1995) and reported
on the left top of figure 3 in the paper by Johansen and Sornette (2000).
The material fitting parameters are $a=-0.101$, $b=-0.131$, $c=-10.86$,
$\alpha=0.839$, and $\omega=3.16$. Again, the rate $f'(x)$ diverges at
the critical point of fracture, with the critical index $\Delta=-0.161$.

\section{Conclusion}

A general law of energy release for stressed composite materials is
derived, being valid in the whole interval starting from the initially
imposed load till the very moment of fracture. This law is presented by
the scale-invariant dimensionless form (10). The derivation of formula
(10) is based on the extrapolation technique of the self-similar
approximation theory. In the present case, we extrapolate the
log-periodically decorated power law that is asymptotically valid only
in the vicinity of materials rupture. A specific feature of the present
extrapolation is the occurrence of complex-valued exponents.

The energy-release law (10) is compared with experimental data by fitting
the material parameters. For this purpose, we accomplished experiments
with glass-polyester composite plates. Also, we compared the form (10)
with the acoustic-emission measurements for other materials. Correlating
the results for different samples, we notice that more disordered materials
are characterized by larger values of the log-frequency $\omega$ and by
smaller absolute values of the critical index $\Delta=\alpha-1$. This index
is negative, which reflects the divergence of the energy release rate at the
moment of rupture. These observations confirm that the fracturing process
is analogous to a critical phenomenon, and the point of rupture is somewhat
equivalent to a critical point.

The knowledge of the general law of energy release for stressed composite
materials not only clarifies the qualitative physical features of the
fracturing process but can also be useful for a quantitative description
of the behaviour of such materials and, hopefully, even for predicting
the global failure of the latter. The possibility of predicting the time
of global materials failure would be of great importance for industrial
applications.

\newpage

{\Large{\bf Appendix}}

\vskip 5mm

Here we give a brief account of the basic ideas of the self-similar
approximation theory, employed in this paper for deriving the general law
of energy release. We shall stress only the principal points, while all
technicalities can be found in the cited references.

Suppose we aim at finding a function $f(x)$ whose exact form is not
known, but for which we can get a sequence $\{ f_k(x)\}$ of approximate
expressions numbered by $k=0,1,2,\ldots$. For the time being, the
physical nature of the function $f(x)$ is of no importance. The approach
is very general and can be applied to arbitrary functions. The sequence
$\{ f_k(x)\}$ can be obtained by a kind of perturbation theory or by an
iterative procedure. The standard situation in physical applications is
that the sequence $\{ f_k(x)\}$ either very poorely converges or, as in
the majority of cases, fastly diverges. For example, the approximants
$f_k(x)$ can be valid only for asymptotically small $x\rightarrow 0$,
which happens for the expansion (6), but have no sense for the finite
values of $x$. How could one extrapolate $f_k(x)$ to the region of their
finite variables?

The first pivotal idea is the introduction of {\it control functions}
(Yukalov, 1976). Let us reorganize the sequence $\{ f_k(x)\}$ to another
sequence $\{ F_k(x,u)\}$ of the approximants
$$
F_k(x,u) ={\cal R}\{ f_k(x)\}
$$
by means of a transformation ${\cal R}$ introducing a set of trial
parameters $u$. The transformation ${\cal R}$ is assumed to have an
inverse ${\cal R}^{-1}$, such that
$$
{\cal R}^{-1}\{ F_k(x,u)\} = f_k(x) \; .
$$
Then, in each approximation order $k$, the trial parameters $u$ are to be
replaced by the control functions $u_k(x)$, yielding $F_k(x,u_k(x))$. The
control functions are chosen so that to make convergent the sequence
$\{\tilde f_k(x)\}$ of the {\it optimized approximants}
$$
\tilde f_k(x) \equiv {\cal R}^{-1}\{ F_k(x,u_k(x)\} \; . \qquad \qquad
\qquad (A.1)
$$
Note that, if necessary, some part of control functions can be defined
not before the inverse transformation ${\cal R}^{-1}$ but after it, in
order that the optimized approximants would possess the desired symmetry
properties or satisfy boundary conditions. To this end, one may separate
the trial parameters into two groups, say $u$ and $s$. Then one defines a
transformation ${\cal R}_s$ giving
$$
F_k(x,u,s)={\cal R}_s\{ f_k(x)\} \; .
$$
Now, generalizing the form (A.1), the optimized approximants are
specified as
$$
\tilde f_k(x)\lim_{s\rightarrow s_k(x)}{\cal R}^{-1}\{ F_k(x,u_k(x),s)\}
\; . \qquad \qquad \qquad (A.2)
$$
The method of obtaining a convergent sequence $\{\tilde f_k(x)\}$ of the
optimized approximants (A.1) or (A.2) is called the {\it optimized
perturbation theory}. This approach is nowadays widely used in various
applications (see survey in Yukalov and Yukalova, 2002).

In order to improve further the accuracy of calculations and to obtain an
effective tool for controlling the convergence of the approximation
sequences, it has been necessary to generalize the approach based on the
usage of control functions. The new principal idea has been to interpret
the passage from one successive approximation to another as the motion on
the manifold of approximants, with the approximation order playing the
role of discrete time. This motion can be formalized by means of {\it group
self-similarity} (Yukalov, 1990, 1991, 1992). With this aim in view, we
need to consider again the transformed sequence $\{ F_k(x,u)\}$. Defining
the function $x_k(\varphi)$ by the {\it reonomic constraint}
$$
F_0(x,u_k(x)) = \varphi\; , \qquad x=x_k(\varphi) \; ,
$$
we introduce
$$
y_k(\varphi) \equiv F_k(x_k(\varphi),u_k(x_k(\varphi))) \; .
\qquad \qquad \qquad (A.3)
$$
Conversely, from $y_k(\varphi)$ we can return back to
$$
F_k(x,u_k(x)) = y_k(F_0(x,u_k(x))) \; .
$$
All functions $y_k(\varphi)$, with $k=0,1,2,\ldots$, pertain to a linear
normed space, which is complete in the sense of the norm convergence. We
shall denote this Banach space by ${\cal B}$. The transformation
$y_k:{\cal B}\rightarrow{\cal B}$ is an endomorphism of ${\cal B}$, with
a unitary element given by $y_0(\varphi)=\varphi$. By construction, the
sequences $\{ y_k(\varphi)\}$ and $\{ F_k(x,u_k(x))\}$ are bijective. The
motion in ${\cal B}$ is presentable as the property of group self-similarity
$$
y_{k+p}(\varphi) = y_k(y_p(\varphi)) \; ,
\qquad \qquad \qquad (A.4)
$$
which is a necessary condition for the fastest convergence of the  sequence
$\{ y_k(\varphi)\}$. The semigroup property (A.4) defines a dynamical
system in discrete time, which is termed the approximation cascade. The
cascade velocity is
$$
v_k(\varphi) = y_k(\varphi) - y_{k-1}(\varphi) \; .
$$
The self-similar approximation theory has its name being based on the group
self-similarity (A.4). The fixed point $y^*(\varphi)$ of the approximation
cascade $\{ y_k|k=0,1,2\ldots\}$ corresponds to the sought function $f(x)$.
An approximate fixed point $y_k^*(\varphi)$ is called a quasifixed point.
Embedding the cascade $\{ y_k\}$ into a flow and integrating the evolution
equation, we get the {\it evolution integral}
$$
\int_{y^*_{k-1}}^{y^*_k} \; \frac{d\varphi}{v_k(\varphi)} = \tau_k \; ,
$$
in which $\tau_k$ is a control time required to rich the quasifixed point
$y_k^*(\varphi)$. If the latter is found, then
$$
F_k^*(x,u_k(x)) \equiv y_k^*(F_0(x,u_k(x)))
$$
is also known. From here, we obtain the self-similar approximant
$$
f_k^*(x) \equiv {\cal R}^{-1}\{ F_k^*(x,u_k(x)) \} \; .
\qquad \qquad \qquad (A.5)
$$
Similarly to the generalization (A.2), we may define a transformation
${\cal R}_s$, which introduces a set $s$ of control functions to be
chosen so that to satisfy some symmetry or boundary conditions. All
procedure of constructing the approximation cascade, whose trajectory
$\{ y_k(\varphi,s)\}$ is bijective to the sequence $\{ F_k(x,u_k(x),s)\}$,
is the same as described above and results in the self-similar approximant:
$$
f_k^*(x) \equiv \lim_{s\rightarrow s_k(x)} \; {\cal R}_s^{-1}
\{ F_k^*(x,u_k(x),s) \} \; . \qquad \qquad \qquad (A.6)
$$
The self-similar approximants (A.5) and (A.6) are essentially more accurate
than the optimized approximants (A.1) and (A.2).

It turned out that among different possible transformations ${\cal R}_s$
a very powerful is that one giving the {\it fractal transforms} (Yukalov
and Gluzman, 1997-a-b, 1998-a; Yukalov, 2000). These are defined as
$$
F_k(x,s) \equiv {\cal R}_s\; f_k(x) \equiv x^s f_k(x) \; .
$$
The inverse transformation is
$$
f_k(x) = {\cal R}_s^{-1}\; F_k(x,s) \equiv x^{-s}\; F_k(x,s) \; .
$$
When applied to an asymptotic approximation
$$
f_k(x) = 1 + \sum_{n=1}^k a_n x^{\alpha_n} \; , \qquad \qquad \qquad
(A.7)
$$
the fractal transformation gives
$$
F_k(x,s) = x^s + \sum_{n=1}^k a_n\; x^{\alpha_n+s} \; .
$$
The sequence $\{ F_k(x,s)\}$ of the fractal transforms enjoys better
convergence properties then the initial sequence $\{ f_k(x)\}$.
Expression (A.7) is the standard form usually arising in different
variants of perturbation theory. This form is asymptotic with respect
to $x\rightarrow 0$ and, as a rule, it has no sense for finite $x$,
as in the case of expansion (7). The order of the powers $\alpha_n$ in
the series (A.7) can, in general, be arbitrary, although usually they
are arranged so that $\alpha_n<\alpha_{n+1}$, when $\alpha_n$ are real
or $|\alpha_n|<|\alpha_{n+1}|$, when $\alpha_n$ are complex. Following
the described steps of the self-similar approximation theory, with the
usage of the fractal transforms, for the approximants (A.6), we get the
recurrence relation
$$
f_k^*(x) =\left\{ [f_{k-1}^*(x)]^{\delta_k} + A_k\; x^{\alpha_k}
\right \}^{1/\delta_k} \; , \qquad \qquad (A.8)
$$
in which $A_k\equiv a_k\delta_k\tau_k$ and $\delta_k\equiv -\alpha_k/s_k$.
The relation (A.8) serves as a source for deriving the final form of the
self-similar approximants. For deriving the self-similar exponential
approximants, employed in the present paper, we notice that the series
(A.7) can be reorganized to the iterative form
$$
f_k(x) = 1 + z_1(1+ z_2(\ldots (1+z_k)) \ldots ) \; ,
\qquad \qquad \qquad (A.9)
$$
where
$$
z_n = z_n(x) = \frac{a_n}{a_{n-1}}\; x^{\alpha_n-\alpha_{n-1}} \; .
$$
Applying $k$ times the described procedure to expression (A.9), with
respect to variables $z_n$, we use the known limit
$$
\lim_{s\rightarrow\pm\infty}\; \left (1 +\frac{\tau}{s}\; z\right )^s =
\exp(\tau z) \; .
$$
As a result, we obtain the self-similar exponential approximants
$$
f_k^*(x) =\exp\left ( c_1\; x^{\nu_1}\; \exp \left ( c_2\; x^{\nu_2}
\ldots \exp(c_k\; x^{\nu_k}) \right ) \ldots \right )  \; ,
\qquad \qquad (A.10)
$$
in which the controllers are
$$
c_n \equiv \frac{a_n}{a_{n-1}}\; \tau_n \; \qquad
\nu_n \equiv \alpha_n-\alpha_{n-1} \; .
$$
This way of deriving the expression (A.10) from an initially given
asymptotic expansion (A.7) has been employed in obtaining the law (1)
from the log-periodic approximation (7). The principal point is that
the expansions (6) or (7) and (A.7) are only asymptotic, being valid
solely for $x\rightarrow 0$; while the laws (10) and (A.10) being the
self-similar extrapolations, are justified for finite values of the
variable $x$.

\newpage

\begin{itemize}

\item
Anifrani J. C., Le Floch C., Sornette D. and Souillard B., Universal
log-periodic correction group scaling for rupture stress prediction from
acoustic emission. J. Phys. I France {\bf 5} (1995) 631-638.

\item
Bernasconi J. and Schneider W. R., Diffusion on a one-dimensional lattice
with random asymmetric transition rates. J. Phys. A: Math. Gen.
{\bf 15} (1983) 729-734.

\item
Borodich F.M., Some fractal models of fracture. J. Mech. Phys. Solids
{\bf 45} (1997) 239-259.

\item
Borodich F.M., Parametric homogeneity and non-classical self-similarity.
Mathematical background. Acta Mechanica {\bf 131} (1998-a) 27-45.

\item
Borodich F.M., Parametric homogeneity and non-classical self-similarity.
Some applications. Acta Mechanica {\bf 131} (1998-b) 47-67.

\item
Borodich F.M. and Galanov B.A., Self-similar problems of elastic contact
for non-convex punches. J. Mech. Phys. Solids {\bf 50} (2002) 2441-2461.

\item
Broberg B., Cracks and Fractures. Cambridge, London (1999).

\item
Dietrich S. and Sornette D., Log-periodic oscillations for biased
diffusion on random lattice. Physica A {\bf 252} (1998) 271-277.

\item
Herrmann H. J. and Roux S., Statistical Models for the Fracture of
Disordered Media. North-Holland, Amsterdam (1990).

\item
Johansen A. and Sornette D., Critical ruptures. Eur. Phys. J. B
{\bf 18} (2000) 163-181.

\item
Kachanov L.M., Introduction to Continuum Damage Mechanics.
Martinus Nijhoff, Dordrecht (1986).

\item
Lamagn\`ere L., Carmona F. and Sornette D., Experimental realization of
critical thermal fuse rupture. Phys. Rev. Lett. {\bf 77} (1996)
2738-2741.

\item
Moura A. and Yukalov V.I., Self-similar extrapolation for the law of
acoustic emission before failure of heterogeneous materials. Int. J.
Fract. {\bf 115} (2002) 3-8.

\item
Moura A., Nechad H., Godin N., El Guerjouma R.  and Baboux J. C.,
Pr\'ediction de la dur\'ee de vie des mat\'eriaux h\'et\'erog\`enes.
Mat\'eriaux 2002 Congress, UBTM, Colloque 4, Tours (2002).

\item
Nauenberg M., Scaling representation for critical phenomena. J. Phys. A
{\bf 8} (1975) 925-928.

\item
Sahimi M. and Arbabi S., Scaling laws for fracture of heterogeneous
materials and rocks. Phys. Rev. Lett. {\bf 77} (1996) 3689-3692.

\item
Sornette D.  and Andersen J. V., Scaling with respect to disorder. Eur.
Phys. J. B {\bf 1} (1998) 353-357.

\item
Sornette D., Discrete scale invariance and complex dimensions. Phys.
Rep. {\bf 297} (1998) 239-270.

\item
Yukalov V.I., Theory of perturbations with a strong interaction. Moscow
Univ. Phys. Bull. {\bf 31} (1976) 10-15.

\item
Yukalov V.I., Statistical mechanics of strongly nonideal systems. Phys.
Rev. A {\bf 42} (1990) 3324-3334.

\item
Yukalov V.I., Method of self-similar approximations. J. Math. Phys.
{\bf 32} (1991) 1235-1239.

\item
Yukalov V.I., Stability conditions for method of self-similar approximations.
J. Math. Phys. {\bf 33} (1992) 3994-4001.

\item
Yukalov V.I. and Gluzman S., Critical indices as limits of control
functions. Phys. Rev. Lett. {\bf 79} (1997-a) 333-336.

\item
Yukalov V.I. and Gluzman S., Self-similar bootstrap of divergent series.
Phys. Rev. E {\bf 55} (1997-b) 6552-6565.

\item
Yukalov V.I. and Gluzman S., Self-similar exponential approximants.
Phys. Rev E {\bf 58} (1998-a) 1359-1382.

\item
Yukalov V.I., Yukalova E.P. and Gluzman S., Self-similar interpolation in
quantum mechanics, Phys. Rev. A {\bf 58} (1998-b) 96-115.

\item
Yukalov V.I.  and Gluzman S., Weighted fixed points in self-similar
analysis of time series. Int. J. Mod. Phys. B {\bf 13} (1999)
1463-1476.

\item
Yukalov V.I., Self-similar extrapolation of asymptotic series and
forecasting for time series. Mod. Phys. Lett. B {\bf 14} (2000)
791-800.

\item
Yukalov V.I., Self-similar approach to market analysis. Eur. Phys.
J. B {\bf 20} (2001) 609-617.

\item
Yukalov V.I. and Yukalova E.P., Self-similar structures and fractal transforms
in approximation theory. Chaos Solitons Fractals {\bf 14} (2002) 839-861.

\end{itemize}

\newpage

\begin{center}
{\large{\bf Figure captions}}

\end{center}

{\bf Figure 1}. Synoptic diagram - Stressed composite material instrumented
with piezoelectric transducers. The electric signals are recorded in real-time
on computer.

\vskip 1cm

{\bf Figure 2}. Dimensionless energy release (10) fitted to our experimental
data.

\vskip 1cm

{\bf Figure 3}. Dimensionless energy release (10), fitted to the
experimental data of Anifrani {\it et al}.

\end{document}